\documentstyle[psfig]{cici}

\def\simg{\mathrel{%
      \rlap{\raise 0.511ex \hbox{$>$}}{\lower 0.511ex \hbox{$\sim$}}}}
\def\siml{\mathrel{%
      \rlap{\raise 0.511ex \hbox{$<$}}{\lower 0.511ex \hbox{$\sim$}}}}
\def\etal{et al$.$ } \def\eg{e$.$g$.$ } \def\ie{i$.$e$.$ } 

\def\reference{\bibitem}

\begin{document}

\title[ X-ray chromatic breaks ]
       { Evidence for chromatic X-ray light-curve breaks in Swift GRB afterglows and their theoretical implications }

\author[Panaitescu \etal ]{\Large  A. Panaitescu$^1$, P. M\'esz\'aros$^{2,3}$, D. Burrows$^3$, J. Nousek$^3$, N. Gehrels$^4$,
                            P. O'Brien$^5$, R. Willingale$^5$ \\
$^1$ Space Science and Applications, Los Alamos National Laboratory, Los Alamos, NM 87545, USA \\
$^2$ Department of Astronomy and Astrophysics, Pennsylvania State University, University Park, PA 16802, USA \\
$^3$ Department of Physics, Pennsylvania State University, University Park, PA 16802, USA \\
$^4$ NASA/Goddard Space Flight Center, Greenbelt, MD 20771, USA  \\
$^5$ Department of Physics and Astronomy, University of Leicester, Leicester, LE 1 7RH, UK }

\maketitle

\begin{abstract}
\begin{small}
 The power-law decay of the X-ray emission of GRB afterglows 050319, 050401, 050607, 050713A, 050802 
and 050922C exhibits a steepening at about 1--4 hours after the burst which, surprisingly, is not 
accompanied by a break in the optical emission. 
 If it is assumed that both the optical and X-ray afterglows arise from the same outflow then, in 
the framework of the standard forward shock model, the chromaticity of the X-ray light-curve breaks 
indicates that they do not arise solely from a mechanism related to the outflow dynamics (\eg energy 
injection) or the angular distribution of the blast-wave kinetic energy (structured outflows or jets). 
 The lack of a spectral evolution accompanying the X-ray light-curve breaks shows that 
these breaks do not arise from the passage of a spectral break (\eg the cooling frequency) either. 
 Under these circumstances, the decoupling of the X-ray and optical decays requires that the 
microphysical parameters for the electron and magnetic energies in the forward shock evolve in time, 
whether the X-ray afterglow is synchrotron or inverse-Compton emission. 
 For a steady evolution of these parameters with the Lorentz factor of the forward shock and an X-ray 
light-curve break arising from cessation of energy injection into the blast-wave, the optical and X-ray 
properties of the above six Swift afterglows require a circumburst medium with a $r^{-2}$ radial 
stratification, as expected for a massive star origin for long GRBs.
 Alternatively, the chromatic X-ray light-curve breaks may indicate that the optical and X-ray emissions 
arise from different outflows. 
 Neither feature (evolution of microphysical parameters or the different origin of the optical and X-ray 
emissions) were clearly required by pre-Swift afterglows. 
\end{small}
\end{abstract}


\section{Introduction}

\begin{figure*}
\centerline{\psfig{figure=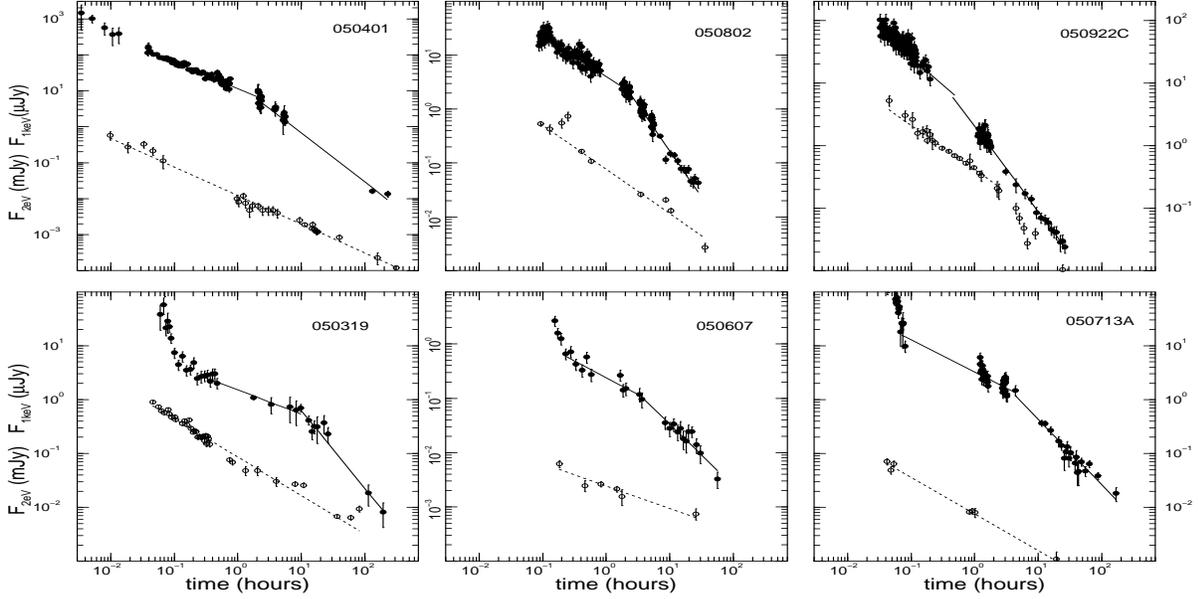,height=8cm,width=16cm}}
\caption{  
   Light-curves of six Swift GRB afterglows showing a chromatic X-ray break which is not seen in the optical
   at the same time.
   Optical data are shown with open symbols and are fit with a power-law decay (dotted lines).
   X-ray data are shown with filled symbols are fit with a broken power-law (solid lines).
   Optical measurements are from Data are from 
   Wo\'zniak \etal (2005), GCNs 3120 (T. Yoshioka), 3124/3140 (D. Sharpov) (050319);
   De Pasquale \etal (2006), Watson \etal (2006), Rykoff \etal (2005) (050401);
   GCNs 3531 \& 3540 (J. Rhoads) (050607);
   Guetta \etal (2006) (050713A);
   GCNs 3739/3745 (K. McGowan), 3744 (E. Pavlenko), 3765 (V. Testa) (050802);
   GCNs 4012 (E. Rykoff), 4015 (P. Jakobsson), 4016 (E. Ofek), 4023 (D. Durig), 4026 (T. Henych), 
   4041 (S. Hunsberger), 4046 (S. Covino), 4048 (M. Andreev), 4040 (J. Fynbo), 4095 (W. Li) (050922C).  
  }
\end{figure*}

\begin{table*}
 \caption{X-ray and optical properties for the afterglows of Figure 1}
\begin{tabular}{cccccccccccc}
 \hline
   GRB   & $\alpha_{x1}$ & $t_b$ & $\alpha_{x2}$ & $\beta_x$     & Refs. &  $\alpha_o$   & $t_o/t_b$ &       s       &  e  &  b  &  i   \\
         &     (i)       &  (ii) &     (iii)     &     (iv)      &       &     (v)       &    (vi)   &     (vii)     &(vii)& (ix)& (x)  \\
 \hline
  050319 & $0.54\pm0.04$ &  7.5  & $1.14\pm0.20$ & $0.75\pm0.05$ &  (1)  & $0.71\pm0.02$ & 0.006--10 & $2.28\pm0.04$ & 3.8 & 5.4 & -3.6 \\
  050401 & $0.65\pm0.02$ &  1.2  & $1.39\pm0.05$ & $1.00\pm0.13$ & (2,3) & $0.80\pm0.03$ & 0.01--250 & $2.17\pm0.08$ & 5.6 & 5.5 & -1.7 \\
  050607 & $0.61\pm0.11$ &  3.3  & $1.12\pm0.07$ & $1.15\pm0.11$ &  (4)  & $0.41\pm0.10$ & 0.05--7   & $1.87\pm0.08$ & 3.6 & 4.5 &  0.0 \\
  050713A& $1.02\pm0.07$ &  4.4  & $1.45\pm0.06$ & $1.07\pm0.04$ &  (5)  & $0.68\pm0.05$ & 0.008--4  & $2.06\pm0.04$ & 1.9 & 5.8 & -1.4 \\
  050802 & $0.64\pm0.10$ &  1.7  & $1.66\pm0.06$ & $0.91\pm0.19$ &  (6)  & $0.82\pm0.03$ & 0.06--20  & $2.24\pm0.13$ & 7.2 & 7.6 & -4.1 \\
  050922C& $0.80\pm0.10$ &  0.3  & $1.19\pm0.02$ & $1.10\pm0.09$ &  (6)  & $0.71\pm0.05$ & 0.1--10   & $2.06\pm0.06$ & 2.4 & 4.2 & -0.3 \\
 \hline
\end{tabular}
\begin{minipage}{180mm}
 (i) pre-break X-ray decay index; (ii) epoch of X-ray light-curve break, in hours; (iii) post-break X-ray decay index; 
 (iv) slope of X-ray spectrum; (v) optical decay index (same before and after $t_b$); (vi) time-range of optical power-law decay relative 
 to the X-ray break epoch; (vii)--(x) parameters for the medium structure (equation \ref{n}) and evolution of blast-wave energy \& 
 microphysical parameters (equation \ref{expo}).
 References for (i)--(iv): (1) Cusumano \etal (2006), (2) Watson \etal (2006), (3) De Pasquale \etal (2006), (4) Nousek \etal (2006), 
  (5) Morris \etal (2006), (6) O'Brien \etal (2006). 
\end{minipage}
\end{table*}

 The standard blast-wave model for GRB afterglows (\eg M\'esz\'aros \& Rees 1997), where the afterglow
emission arises from the external shock which dissipates GRB ejecta kinetic energy, energizes the 
circumburst medium, accelerates relativistic electrons and generates magnetic fields, producing
synchrotron emission, has been largely confirmed by the numerous X-ray, optical, and radio afterglows
discovered so far. Their light-curves exhibited a power-law decay, as expected from the power-law
decay of the shock Lorentz factor with observer time and the power-law distribution with energy of 
shock-accelerated electrons (together, these two properties lead to synchrotron spectrum characteristics 
with a power-law temporal evolution). In general, the optical and X-ray light-curve decay indices were 
found consistent with the slope of the afterglow spectrum (\eg Wijers, Rees \& M\'esz\'aros 1997).

 Further testing of the standard blast-wave model can be done by comparing the light-curve decays at
different wavelengths. For instance, the light-curve breaks (\ie steepening of power-law decay) arising
from the dynamics or the collimation of the GRB outflow (Rhoads 1999) should be achromatic, occurring 
simultaneously at all frequencies. The lack of a sufficiently extended X-ray monitoring of pre-Swift 
afterglows prevented us to test if the breaks observed in the optical light-curves of many GRB afterglows 
were achromatic, with the possible exceptions of GRBs 010222 and 030329. For the former, a single 10 day
X-ray measurement (in't Zand \etal 2001) indicates the existence of a break in the X-ray light-curve 
at $\simg 2$ days, which could have been simultaneous with the gradual optical break. For the latter, 
a break in the X-ray light-curve at 0.5 day (Tiengo \etal 2003) is accompanied by a steepening of the 
optical light-curve decay (Lipkin \etal 2004), however the variability of the optical emission makes
a jet-break identification difficult.

 In the last year, the Swift satellite has monitored the X-ray emission of dozens of afterglows, and
has shown that, after a rather shallow decay, the X-ray light-curves exhibit a steepening at about
1--4 hours after the burst. Because of their dimness, only a small fraction of these afterglows were followed
in the optical. The only cases so far for an X-ray light-curve break possibly accompanied by an optical 
steepening are the GRB afterglows 050525A (Blustin \etal 2006) and 050801 (Rykoff \etal 2006). 
The X-ray light-curve of the former appears more complex than just a broken power, with an uncertain 
break time, while the optical emission shows a brightening (Klotz \etal 2005) around the epoch of the
X-ray break. The optical and X-ray breaks of the latter occur much earlier than for other afterglows, 
at only 4--5 minutes after the burst.
 
 Surprisingly, as shown in Figure 1, the optical light-curves of other GRB afterglows do not exhibit a
steepening at the epoch of the X-ray break. Summarizing their temporal properties, the X-ray light-curves 
decays steepen from $F_x \propto t^{-0.8\pm0.2}$ to $F_x \propto t^{-1.4\pm0.2}$ at $t_b = 1-4$ hours, 
however the optical light-curves, which start 1--2 decades before $t_b$, maintain a $F_o \propto 
t^{-0.6\pm0.2}$ decay after $t_b$. The evidence for the X-ray break chromaticity is weak for GRBs 050607 
and 050713A (lower panel of Figure 1), for which there is only one post-break optical measurement 
(possibly contaminated by the host galaxy emission) but is rather compelling for GRBs 050401, 050802 
and 050922C (upper panel of Figure 1). In particular, no optical light-curve break is seen for the GRB 
afterglow 050401 for 2.5 decades in time after $t_b$ (Watson \etal 2006). Only the optical emission of 
the GRB afterglow 0509022C exhibits a steepening, but at $10\,t_b$. 

 The purpose of this article is to investigate the conditions required for the external shock to produce 
such a chromatic X-ray light-curve break. The afterglow emission is described by
\begin{equation}
 F_\nu (t) \propto \nu^{-\beta} t^{-\alpha} 
\end{equation}
as expected in the standard blast-wave model for GRB afterglows (\eg Paczy\'nski \& Rhoads 1993, 
M\'esz\'aros \& Rees 1997, Sari, Narayan \& Piran 1998), with "o" and "x" for the subscript "$\nu$" 
designating optical and X-ray quantities, respectively. Additionally, the pre- and post-break X-ray 
light-curve decay indices are denoted by $\alpha_{x1}$ and $\alpha_{x2}$.

\section{Model interpretation}

 The lack of an optical light-curve break contemporaneous with that seen in the X-ray shows that 
the X-ray light-curve breaks do not arise from the outflow collimation (a jet) or angular structure
(where the ejecta kinetic energy is a function of direction). For this reason, we consider only 
the case of an uniform outflow which is sufficiently relativistic that its boundaries are not yet 
visible to the observer.

\subsection{Cooling frequency crossing the X-ray}

 The possibility that the X-ray light-curve break is caused by the passage of a spectral break through 
the X-rays is largely excluded by the lack of a significant spectral evolution (Cusumano \etal 2006, 
De Pasquale \etal 2006, Nousek \etal 2006, Morris \etal 2006), a fact also pointed out by Fan \& Piran 
(2006) for the X-ray afterglows 050319 and 050401. In contrast, if the cooling break frequency, $\nu_c$, 
(which is the frequency at which radiate the electrons whose cooling timescale is equal to the dynamical 
timescale) had crossed the X-ray band, then the X-ray continuum should have softened by $\Delta \beta_x = 
0.5$ for a homogeneous circumburst medium (for which $\nu_c \propto t^{-1/2}$) or hardened by $\Delta 
\beta_x = -0.5$ for a wind-like medium (for which $\nu_c \propto t^{1/2}$).

 Should the above spectral evolution be measured across $t_b$ in other X-ray afterglows, the passage of 
$\nu_c$ through X-ray domain would lead to a break of the X-ray light-curve with the following properties:
\begin{equation}
 ({\rm homogeneous\; medium})\quad  \alpha_{x1} = \alpha_o \;, \alpha_{x2} = \alpha_o + 1/4 
\end{equation} 
\begin{equation}
 ({\rm wind\; medium})\quad  \alpha_{x1} = \alpha_o - 1/4 \;, \alpha_{x2} = \alpha_o 
\end{equation} 
if the electron radiative cooling is synchrotron-dominated (\ie Compton parameter $Y<1$). In this case,
the magnitude of the X-ray light-curve break is $\Delta \alpha_x \equiv \alpha_{x2} - \alpha_{x1} = 1/4$, 
\ie smaller than that observed for most (if not all) Swift afterglows.

 If inverse-Compton scattering dominated the electron cooling ($Y>1$) then, for a homogeneous medium, 
the cooling frequency would decrease more slowly or can even increase, which leads to an even smaller 
X-ray break magnitude: $\Delta \alpha_x < 1/4$ (Panaitescu \& Kumar 2001). 
A stronger X-ray break is produced by a wind-like medium, where an inverse-Compton dominated electron 
cooling accelerates the increase in time of $\nu_c$, leading to $1/4 < \Delta \alpha_x \leq 5/4$ 
(Panaitescu \& Kumar 2001). However, in this case, the optical and post-break X-ray decay indices should
be equal ($\alpha_{x2} = \alpha_o$), in contradiction with what is observed for all afterglows in Table 1. 

 Therefore, the passage of the cooling frequency through the X-ray as a reason for the X-ray light-curve
breaks shown in Figure 1 is incompatible not only with the lack of a X-ray spectral evolution but also
with the measured optical and X-ray decay indices.

\subsection{Synchrotron emission in the X-ray}

 In a model where both the optical and X-ray emissions arise from the same afterglow synchrotron component,
the optical and X-ray light-curves can have different behaviours if the cooling frequency lies between the 
optical and X-ray. Then one can determine its evolution from the optical and X-ray decay indices. 
In the external shock model, the optical and X-ray afterglow fluxes are related by
\begin{equation}
  F_x = F_o \left( \frac{\nu_o}{\nu_c} \right)^{\beta_o} 
            \left( \frac{\nu_c}{\nu_x} \right)^{\beta_x} \;.
\label{fxfo}
\end{equation}
It follows that $\nu_c$ must have the following evolution
\begin{equation}
 \frac{d\ln \nu_c}{d\ln t} = \frac{\alpha_o - \alpha_x}{\beta_x -\beta_o}  = 
            2 (\alpha_o - \alpha_x) 
\label{nub}
\end{equation}
where $\beta_x = \beta_o + 1/2$ (for $\nu_o < \nu_c < \nu_x$) was used.

 For the afterglows of Table 1, $\alpha_{x2} -\alpha_o \in (0.4,0.9)$, implying a post-break decrease of the 
cooling frequency, $\nu_c \propto t^{-1.3\pm0.5}$, which is much faster than that expected in the standard 
forward shock model. In this model, for constant microphysical parameters and no energy injection in the
blast-wave, the steepest decrease is obtained for a homogeneous medium and synchrotron-dominated electron 
cooling: $\nu_c \propto t^{-1/2}$. Energy injection in the forward shock by means of some GRB ejecta 
catching-up with it leads to a faster decrease of $\nu_c$ but cannot by itself produce a chromatic X-ray 
light-curve break because it alters the afterglow flux below the cooling frequency as well. Instead, the 
fast decrease of $\nu_c$ suggests that the fraction of the post-shock energy in magnetic fields, $\epsilon_B$, 
is evolving. Because the afterglow emission also depends on the fraction $\epsilon_i$ of the post-shock 
energy that is imparted to electrons, we also allow it to be time-dependent.

 Given that the afterglow light-curves are power-laws in observer time, evolution of the microphysical 
parameters and blast-wave kinetic energy $E$ (allowing for energy injection) must also be power-laws in $t$. 
The latter implies that the blast-wave Lorentz factor $\Gamma$ is also a power-law in time, hence $E$, 
$\epsilon_B$, and $\epsilon_i$ must evolve as power-laws of $\Gamma$:
\begin{equation}
 E (> \Gamma) \propto \Gamma^{-e} \;,\;\;  \epsilon_B \propto \Gamma^{-b} \;,\;\; \epsilon_i \propto \Gamma^{-i} 
\label{expo}
\end{equation}
where $E(>\Gamma)$ denotes the energy of all ejecta with Lorentz factor larger than a given $\Gamma$ which
have caught-up with the forward shock. For a decelerating blast-wave, the increase of its kinetic energy 
is equivalent to $e > 0$. We also allow for a power-law stratification of the ambient medium,
\begin{equation}
 n(r) \propto r^{-s}  \quad (s < 3) 
\label{n}
\end{equation}
to be determined from the optical and X-ray afterglow properties. 
The condition $s < 3$ ensures that the blast-wave is decelerated, for any $e > 0$. 

 The synchrotron optical ($\nu_o < \nu_c$) and X-ray ($\nu_c < \nu_x$) light-curves can be inferred from
the decay indices given in Panaitescu \& Kumar (2004) by using
\begin{equation}
 \frac{d\ln \Gamma}{d\ln t} = - \frac{3-s}{e+8-2s} 
\label{G}
\end{equation}
which follows from assuming an adiabatic blast-wave, $\Gamma^2 M \propto E$, where $M \propto r^{3-s}$ is the
mass of the swept-up medium, and from the relation between the blast-wave radius and photon arrival time,
$R \propto \Gamma^2 t$. The results are
\begin{equation}
 \alpha_o = \frac{s}{8-2s} + \frac{3p-3}{4} - \frac{3-s}{e+8-2s} \times
\label{aosy}
\end{equation}
\begin{displaymath}
    \quad \quad \quad \left[ \left( \frac{p+3}{4} - \frac{s}{8-2s} \right) e + (p-1)i + \frac{p+1}{4}b \right] 
\end{displaymath}
\begin{equation}
 \alpha_x = \frac{3p-2}{4} - \frac{3-s}{e+8-2s} \left[ \frac{p+2}{4} e + (p-1)i + \frac{p-2}{4}b \right] 
\label{axsy}
\end{equation}
where $p = 2 \beta_x$ is the exponent of the power-law electron distribution with energy in the forward shock:
\begin{equation}
 \frac{dN}{d\gamma} (\gamma > \gamma_i) \propto \gamma^{-p} 
\end{equation}
$\gamma_i$ being the typical (comoving) Lorentz factor of the shock-accelerated electrons.

 We do not consider a scenario where the evolutions of the microphysical parameters change at $t_b$ to yield
an X-ray light-curve break because such a scenario is very contrived: those changes in the evolution of 
$\epsilon_i$ and $\epsilon_B$ would have to "conspire" to leave the optical power-law light-curve unaffected. 
A more natural scenario is that where the evolution of the microphysical parameters with the blast-wave Lorentz
factor is steady (\ie constant $i$ and $b$ in equation \ref{expo}) and the X-ray light-curve break is caused 
by a change in the evolution of $\Gamma$. The latter could be due to $i)$ the blast-wave encountering the 
termination shock of the burst progenitor's wind or $ii)$ cessation of energy injection into the blast-wave.

 In the former scenario, the termination shock marks the transition between the freely expanding wind of a
massive star and the homogenized shocked wind. Substituting $s=2$ at $t < t_b$ and $s=0$ at $t > t_b$ in
equation (\ref{axsy}), one can find the parameter $e$ for the energy injection. However, for all the afterglows
listed in Table 1, we obtain an unphysical result: $e < 0$, therefore this scenario does not work. 

 In the second scenario, the cessation of energy injection implies a lower cut-off to the Lorentz factor
of the incoming ejecta. From equation (\ref{aosy}) it can be shown that the condition that the optical decay 
index remains unchanged when $e(t<t_b) >0$ switches to $e(t>t_b)=0$ leads to
\begin{equation}
  (p-1)i + \frac{p+1}{4}b = 2(p+3) - \frac{s}{2} (p+5) \;.
\label{pibs}
\end{equation}
There is no obvious reason for the evolution of the microphysical parameters with the blast-wave Lorentz factor
to "know" about the structure of the ambient medium, hence this scenario remains contrived. The above condition
also shows why the standard forward shock model with energy injection cannot explain the chromaticity of the
X-ray light-curve break: for $i=b=0$ (\ie constant microphysical parameters), the resulting medium structural 
parameter, $s = 4(p+3)/(p+5)$, leads to an optical decay index $\alpha_o = 2\beta_x$ which, for the X-ray 
spectral slopes given in Table 1, is much steeper than observed.

 Substituting equation (\ref{pibs}) in (\ref{aosy}), the optical decay index requires a circumburst 
medium with 
\begin{equation}
   s = 4\; \frac{\alpha_o+3}{p+5} \;.
\label{s}
\label{sbx}
\end{equation}
As shown in Table 1, the optical decay index $\alpha_o$ and X-ray spectral slope $\beta_x$ of the six afterglows
with chromatic X-ray breaks require a circumburst medium with a wind-like stratification. For $s\simeq 2$ and 
$p = 2\beta_x \simeq 2.0$, equation (\ref{pibs}), which is the condition for the lack of an optical break, 
becomes 
\begin{equation}
 \frac{1}{3}\, i + \frac{1}{4}\, b = 1 \;.
\label{ib}
\end{equation}
The energy injection parameter $e$ is determined by the steepening of the X-ray light-curve decay.
From equation (\ref{axsy}), we obtain 
\begin{equation}
  e = \frac{(8-2s) (\alpha_{x2} - \alpha_{x1})}{\alpha_{x1} + 2 - 0.25\, s (p+2)} \;.
\label{e}
\end{equation}
Finally, the parameters $i$ and $b$ can be determined from the decays of the optical and post-break X-ray
light-curves. 
Transforming to observer time with the aid of equation (\ref{G}), the average evolutions given in Table 1 are
\begin{equation}
 t < t_b \;:\;  \Gamma \propto t^{-0.12\pm0.04} \;,\;
             \epsilon_B \propto t^{0.6\pm0.2}   \;,\; \epsilon_i \propto t^{-0.2\pm0.1} 
\end{equation}
\begin{equation}
 t > t_b \;:\;  \Gamma \propto t^{-0.24\pm0.02} \;,\;
             \epsilon_B \propto t^{1.3\pm0.2}   \;,\; \epsilon_i \propto t^{-0.4\pm0.4} 
\end{equation}
and $E \propto t^{0.4\pm0.1}$ before the X-ray break epoch.

\subsection{Inverse-Compton emission in the X-ray}

 If the circumburst medium is sufficiently dense then inverse-Compton (iC) emission can outshine synchrotron
in the X-rays. Alternatively, the synchrotron spectrum could have a cut-off below the X-ray, as the shock-accelerated 
electrons may not acquire sufficient energy (Li \& Waxman 2006). We consider here the case where the optical 
afterglow is synchrotron emission and the X-ray afterglow arises from inverse-Compton scattering.

 The iC continuum has spectral breaks at the frequencies $\gamma_i^2 \nu_i$ and $\gamma_c^2 \nu_c$, where 
$\nu_i$ is the frequency at which the $\gamma_i$-electrons radiate and $\gamma_c$ is the Lorentz factor 
of the electrons which cool radiatively on a timescale equal to the dynamical timescale (\ie they radiate 
at the cooling frequency $\nu_c$). In the standard model for the forward shock emission (\eg Sari \etal 1998,
Panaitescu \& Kumar 2001), it can be shown that the decay indices of the iC light-curves are: 
\begin{equation}
 \alpha_x = \frac{(18-5s)p+c_1(s)}{16-4s} - \frac{3-s}{e+8-2s} \times
\label{ax}
\end{equation}
\begin{displaymath}
 \quad \quad \quad  \left[ \frac{(6-s)p+c_2(s)}{16-4s}\, e + c_3(p)\, i + c_4(p)\, b \right]  
\end{displaymath}
where
\begin{equation}
 c_1 = 11s - 22 ,\; c_2 = 14 - 9s ,\; c_3 = 2p-2 ,\; c_4 = \frac{p+1}{4}
\end{equation}
for $\gamma_i^2 \nu_i < \nu_x < \gamma_c^2 \nu_c$ (in which case $p = 2\beta_x+1$) and
\begin{equation}
 c_1 = 6s - 20 ,\; c_2 = 4 - 2s ,\; c_3 = 2p-2 ,\; c_4 = \frac{p-6}{4}
\end{equation}
for $\nu_x > \max\{\gamma_i^2 \nu_i, \gamma_c^2 \nu_c\}$ (in which case $p = 2\beta_x$).
Synchrotron-dominated electron cooling ($Y < 1$) was assumed for the latter expressions of $c_3$ and $c_4$;
if iC dominates ($Y > 1$) then $c_3 = (2p-3)$ and $c_4 = (p-2)b/4$.

 Aside from the above two possible locations of the X-ray domain relative to the iC spectral breaks, the cooling 
frequency could be either above or below the optical range (the optical light-curve decay index being given 
by equations \ref{aosy} and \ref{axsy}, respectively), which yields four possible combinations to be considered. 
Just as for the case where the X-ray afterglow were attributed to synchrotron emission, we obtain an unphysical 
solution ($e < 0$) if the X-ray light-curve break is attributed to the blast-wave encountering the wind termination 
shock. For the scenario where this break arises from cessation of energy injection, $e > 0$ solutions
exist only for $\nu > \max\{\gamma_i^2 \nu_i, \gamma_c^2 \nu_c\}$ and, for three afterglows, only if the cooling 
frequency is higher than optical frequencies (in this case, the lack of an optical light-curve break leads to 
equation \ref{pibs}). 
The magnitude of the X-ray light-curve break determines the energy injection parameter $e$ and the 
result is the same as when the X-ray afterglow was assumed to be synchrotron emission (equation \ref{e}). 
However, the evolution of the microphysical parameters with the blast-wave Lorentz are weaker now because the 
decay indices of the iC light-curves have a stronger dependence on $\epsilon_i$ and $\epsilon_B$.

\section{Discussion}

 The lack of an optical light-curve break contemporaneous with the steepening of the power-law decay observed 
at $1--4$ hours in the X-ray light-curves of several Swift afterglows (Figure 1) is a puzzling new feature. 
It does not originate from the outflow collimation (such breaks should be chromatic) nor from the passage of 
a spectral break through the X-ray (such light-curve breaks should be accompanied by a spectral evolution).
If the optical and X-ray afterglows are the same synchrotron forward-shock emission then the only possibility 
left is that the chromatic X-ray light-curve breaks are caused by the evolution of a spectral break $\nu_b$ 
located between the optical and X-ray. That evolution could be determined from the optical and X-ray decay 
indices and spectral slopes (equation \ref{nub}). However, the optical intrinsic spectral slope may be difficult
to measure because of dust extinction in the host galaxy, as is the case of GRB 050401 (Watson \etal 2006). 

 For the GRB afterglow 050401, whose optical emission has been monitored well after the X-ray light-curve 
break, the required post-break evolution of $\nu_b$ is barely compatible with the lack of a break in the 
optical light-curve until 10 days: in the most favourable situation, where $\nu_b$ was just above the XRT's 
0.3 keV threshold at $t_b = 1.2$ hours, the $\nu_b (t>t_b) \propto t^{-1.25}$ required by the post-break
X-ray decay implies that this spectral break should cross the R-band at 8 days. 
Over the four decades in time during which the optical afterglow of GRB 050401 exhibits an unbroken 
power-law decay, the energy of the electrons radiating in the optical increases by a factor 30, for a 
homogeneous circumburst medium (100 for a wind-like medium). Moreover, over four decades in time, the ratio 
of the energy of the optical-emitting electrons to that of the peak of the electron distribution increases 
by a factor $10^3$. These facts suggest that the spectral break at $\nu_b$ is not associated with a break in 
the distribution with energy of the shock-accelerated electrons, as in this case it would be difficult to 
understand why $\nu_b$ remains above the optical over a $10^4$-fold increase in time.

 If the spectral break $\nu_b$ is the cooling frequency ($\nu_c$) then its evolution required by the decay 
indices listed in Table 1 is $\nu_c (t>t_b) \propto t^{-0.8}$ or faster, exceeding the fastest decay expected 
for the cooling frequency in the standard blast-wave model: $\nu_c \propto t^{-1/2}$, for a homogeneous 
circumburst medium (for a wind-like medium, $\nu_c$ should increase at least as fast as $t^{1/2}$). This 
shows that evolving microphysical parameters for the electron energy and magnetic field are required to 
explain the chromatic X-ray light-curve breaks. This feature of the forward shock model was not previously 
required by the analysis of the broadband emission of pre-Swift afterglows. 

 Attributing the X-ray light-curve break to a sudden change in the evolution of the microphysical parameters
seems ad-hoc and contrived because their evolution would have to be such that a break in the optical light-curve
is not produced. A less contrived scenario is that the microphysical parameters have a steady evolution with 
the blast-wave Lorentz factor and that the X-ray light-curve break arises from cessation of energy injection
in the forward shock (as previously proposed by Nousek \etal 2006, Panaitescu \etal 2006, Zhang \etal 2006).
Then the lack of an optical light-curve break simultaneous with that seen in the X-rays requires that the
evolution of the microphysical parameters satisfies equation (\ref{ib}), whether the X-ray emission is 
synchrotron or inverse-Compton. 

 In this scenario, to accommodate the optical and X-ray properties of the six GRB afterglows considered here 
requires that the stratification of the circumburst medium (Table 1) is that of a stellar wind, $n \propto r^{-2}$, 
as expected from a massive star progenitor for long GRBs (\eg Woosley 1993). 
However, such an ambient medium was not found to be consistent in all cases with the jet-break observed in
optical light-curves. For instance, the sharp optical breaks of GRB afterglows 990510, 
000301C and 011211 are better modelled with a homogeneous external medium, whether the break is attributed to 
seeing the jet boundary or the bright core of a structured outflow seen off-axis (Panaitescu 2005). In a 
wind-like medium, the slower deceleration of the blast-wave leads to a more gradual steepening of the 
power-law decay of the afterglow emission (Kumar \& Panaitescu 2000), as was observed in many, but not all, 
pre-Swift afterglows with optical light-curve breaks. The identification of one or the other kind of circumburst 
medium may have to do with the location of the termination shock of the stellar free wind.
For the innermost locations of this shock (at less than $10^{18}$ cm), the boundary of a narrow jet may become 
visible after the jet has crossed the shock and entered the quasi-homogeneous shell of shocked wind, while for 
outer locations the jet-break occurs in the free wind region. The lack of a jet-break signature in the six Swift 
afterglows analyzed here can be explained if their outflows are not tightly collimated. Furthermore, as discussed
above, the chromaticity of their X-ray light-curve breaks indicates that the afterglow emission occurred in the 
freely expanding stellar wind. 

 A difficulty with the above model is that there is no evident reason for which evolution of the microphysical 
parameters should satisfy equation (\ref{ib}). In the absence of a physical reason for this, an alternative
conclusion would be that, at least in some afterglows, the optical and X-ray emissions arise from different 
outflows. However, such a conclusion is not supported by pre-Swift afterglow observations. For example, 
the optical and X-ray emissions of the twelve afterglows modelled numerically by Panaitescu (2005) can be 
explained with a single outflow. The optical and X-ray continua shown in figures 10 and 11 of Nardini \etal 
(2006) also suggest that, in many cases, they could be the same afterglow spectral component, although 
such a test is complicated by the likely presence of the cooling break between the optical and X-ray
and by the possible dimming and reddening of the optical emission by dust in the host galaxy.

 Further work, both theoretical and observational, is needed to verify if afterglows such as those discussed 
here are indeed weakly collimated outflows interacting with a wind environment and leading to shocks with
evolving microphysical parameters or if the optical and X-ray afterglow emissions are produced by different 
regions of the GRB outflow.

\section*{Acknowledgments}
 A.P.'s research was supported in part by the National Science Foundation under Grant No. PHY99-07949.
 D.B. and J.N. acknowledge support by NASA contract NAS5-00136.

\end{document}